\documentclass[12pt,preprint]{aastex}
\shorttitle{Seeing at Sierra Negra}
\shortauthors{Carrasco et al.}
\begin{document}
\title{\bf Optical Seeing at Sierra Negra}

\author{Esperanza Carrasco, Alberto Carrami\~nana,\\
Jos\'e Luis Avil\'es \& Omar Yam}
\affil{Instituto Nacional de  Astrof\'{\i}sica, \'Optica y Electr\'onica,\\ 
Luis Enrique Erro 1, Tonantzintla, Puebla 72840, M\'exico}
\email{bec@inaoep.mx}


\slugcomment{February 25, 2003}

%---------------%
\begin{abstract}
%---------------%
Optical seeing measurements carried out at Sierra Negra, the site of the Large Millimeter
Telescope, are reported. The site, one of the highest peaks of Central Mexico, offers good 
coverage of Northern and Southern hemispheres and we have undertaken several campaigns to 
investigate the astronomical potential of the site in the optical. Here we report on our 
campaign aimed at establishing the seeing quality of the site. We present data of the first 
three campaigns of optical seeing monitoring covering from February 2000 to May 2002, carried 
out with a Differential Image Motion Monitor. The results clearly indicate a sub-arcsec seeing, 
better statistics during the dry season and no dependence with the time of night. We find
no dependence of our results with the integration time used.
%---------------%
\end{abstract}
%---------------%

%%%%%%%%%%%%%%%%%%%%%%
\section{Introduction}
%%%%%%%%%%%%%%%%%%%%%%
Sierra Negra is an extinct volcano in the State of Puebla, Mexico, located at 
18$^{\circ}$~59$^{'}06^{"}$~N latitude, 97$^{\circ}$~18$^{'}53^{"}$~W longitude and 
an altitude of 4580~m above sea level. The mountain is next to Citlaltepetl, the highest 
peak in Mexico, with just 8~km separating both summits. Sierra Negra is about 100~kms West 
from the coast of Veracruz in the Gulf of Mexico and 300~kms from the Pacific Coast. The 
site, administrated by the Instituto Nacional de Astrof\'{\i}sica, \'Optica y Electr\'onica 
(INAOE), is inside the Pico de Orizaba National Park. Easy access is available via 100~km 
motorway from the city of Puebla followed by a 20~km  access road to the summit. The 
journey from Puebla city takes about two hours. 

In February 1996  Sierra Negra was selected, among more than twenty potential sites, as the 
site of the Large Millimeter Telescope (LMT/GTM), now  under construction. The decision was 
based on its low atmospheric water vapour content, with registered opacities at 240~GHz down
to $\la 0.02$. The LMT/GTM is a 50~m antenna optimised for 1-3~mm observations. First light 
and  science operations are planned for  2005. The LMT/GTM is a bi-national project between 
Mexico and the United States, leaded by INAOE in Mexico and the University of Massachusetts, 
at Amherst, in the USA. 

With the development of the Sierra Negra site, INAOE planned to measure its quality for 
optical observations. Because of its altitude and location the site is intrinsically very 
dry, therefore the conditions are likely to be favorable for near infrared and optical
observations. An unknown property of the site is its optical seeing, a key parameter to 
determine how good an astronomical site is. In the last few decades great efforts have been 
dedicated to the development of 8-10~m class diameter telescopes and for instrumentation 
which requires very precise site characterization. Furthermore, the new generation of 
Extremely Large Telescopes will require selecting sites with very good seeing conditions. 

INAOE decided to undertake a first optical seeing and weather measurements campaign starting 
in February 2000 without basic facilities available. A temporary set-up was prepared for the 
February campaign, which was in fact the first astronomical night-time work performed at the 
site. The second campaign started in October 2000, with better facilities such as 5~m tower, 
a container, a suitable power supply and a place to rest -at about 3000~m above sea level- 
after observing. From  May 2001 we started a routine measurements regime.  Here we present 
the results obtained from February 2000 to May 2002.  Weather data have been taken almost 
continuously from November 2000 up to date. The weather analysis will be reported elsewhere. 

%%%%%%%%%%%%%%%%%%%%%%%%%
\section{The Instrument}
%%%%%%%%%%%%%%%%%%%%%%%%%
The data for this work were taken with a Differential Image Motion Monitor  (DA/IAC DIMM) 
developed  by Vernin \& Mu\~noz-Tu\~non (1995), based on the same physical principle 
as the ESO DIMM (Sarazin \& Roddier 1990) but commercialized by the French company LHESA.
The DIMM principle is to produce twin images of a star with the same telescope via 
two entrance pupils and a wedge. The instrument consists of a 20~cm  Celestron telescope, on 
a very robust equatorial mount, with an intensified CCD camera coupled via an optical fiber 
bundle to the CCD, a Matrox frame grabber board  and a PC. The two $D = 60\rm mm$ apertures, 
separated by a distance $d = 140\rm mm$, are located on a mask attached to the 
telescope entrance pupil. A precisely cut wedge placed over one of the pupils 
deviates the incoming light separating the two star images by approximately 30~arcsec.
The intensified CCD camera and the frame grabber register the relative  position of both 
stellar images after computing the centroid position of each. A statistical seeing value is 
assessed based on the variance of the differential image motion after 200 images are taken. 
The measurement corresponds roughly to a wavelength $\lambda = 0.5~\mu$m, as dictated by the 
response of the system.

%%%% begins comment 1:

Because it is a differential method the technique is, in principle, insensitive to erratic 
motions of the telescope introduced by wind or ground vibrations. Sarazin \& Roddier (1990) 
showed that, assuming a Kolmogorov power-law spectrum for the turbulent cells, the 
longitudinal and transverse variances of the differential motion between 
the images, $\sigma_{l}$ and $\sigma_{t}$, are related to the Fried parameter $r_0$ as:

\begin{eqnarray} 
\sigma{_{l}}^{2} &=& 2\lambda^{2}{r_o}^{-5/3}[0.179D^{-1/3}-0.097 d^{-1/3}]\\
\sigma{_{t}}^{2} &=& 2\lambda^{2}{r_o}^{-5/3}[0.179D^{-1/3}-0.145 d^{-1/3}] .
\end{eqnarray}
Two independent $r_0$ values are obtained which, in principle, should have the same value.
The parameter $r_0$ can be imagined as the telescope diameter that would produce a 
diffraction spot of the same size as that produced by the atmospheric turbulence.  
The seeing is given by $s_{FWHM}= 0.98~(\lambda/r_0)$. These computations are carried 
out internally by the instrument, providing measures of $s \equiv s_{FWHM}$ derived from 
the longitudinal and transverse estimates. The DA/IAC DIMM achieves an accuracy better 
than 0.1" for stars brighter than fourth magnitude with a 10~ms time exposure. 
A reliable seeing measurement is attained within twenty seconds.

%%%%  Ends comment 1

%%%%%%%%%%%%%%%%%%%%%%%%%%%%%%%%%%%%%%%%%%
\section{Observations and Data}
\begin{figure}
\plotone{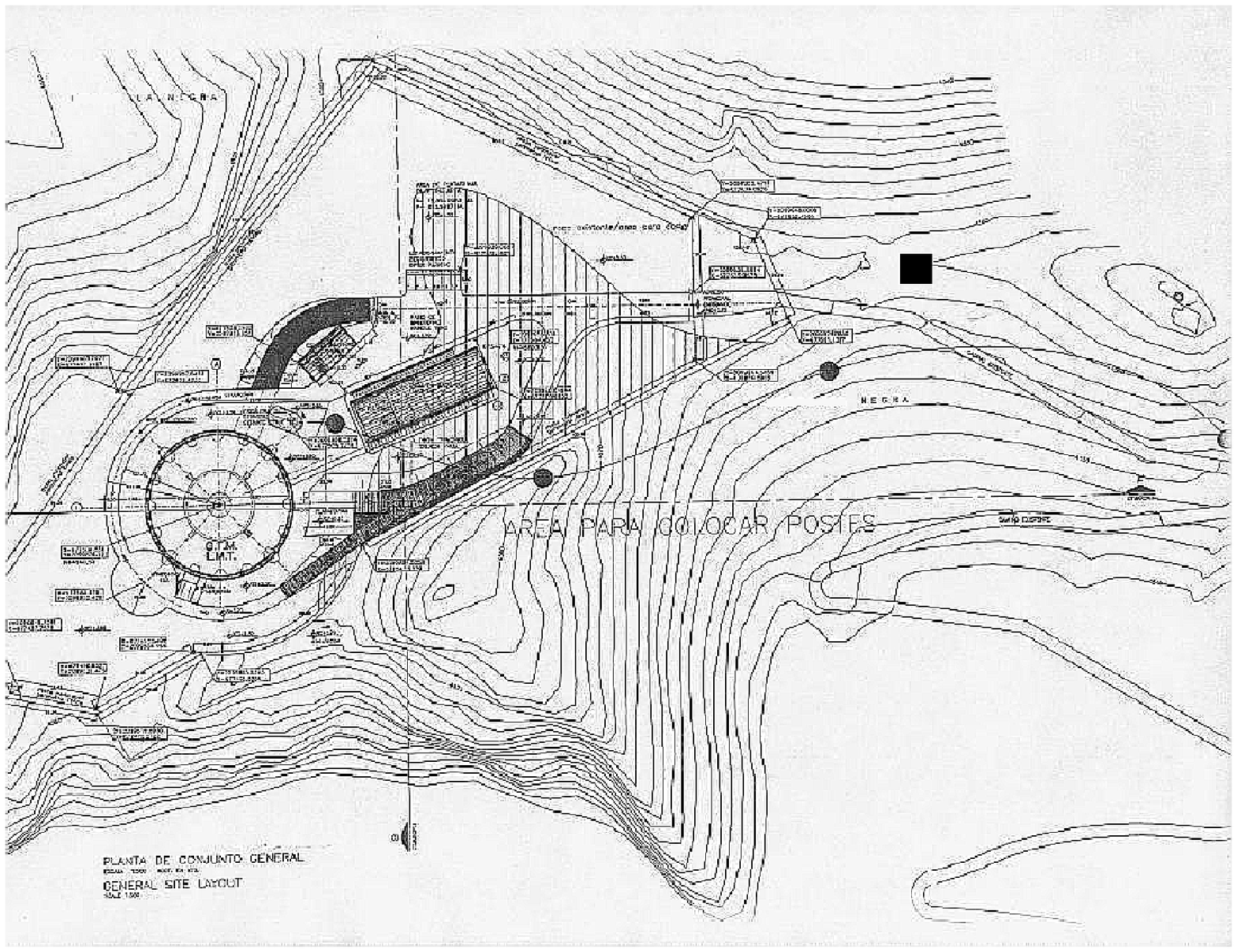}
\caption{Layout of the Large Millimeter Telescope site. North is up and East to 
the right. The LMT track is marked by the 40~m diameter dark circle located at the 
middle left.  The approximate location of the seeing monitor and weather station is 
marked by the black square, East-North-East of the LMT construction. \label{layout}}
\end{figure}

%%%%%%%%%%%%%%%%%%%%%%%%%%%%%%%%%%%%%%%%%%
The seeing monitor location  at the summit is shown in figure \ref{layout}. The LMT is 
located to the  left. The circle corresponding to the LMT track is marked and it has 40 
meters of diameter. The approximate location of the seeing monitor and weather station is 
marked by the black square, to the North-East of the LMT. The  DIMM and the weather station 
are on a five meter height tower  to avoid the surface layer (Vernin \& Mu\~noz-Tu\~non 1994). 
The DIMM is on a tower independent of the platform where the observers move freely without 
affecting the seeing measurement. The tower is near a sharp edge to face directly the 
incoming winds. Thermal equilibrium is ensured by the absence of an enclosure.

The data available cover 85 nights, grouped in three sets:
4 nights between February 22$^{nd}$  and  April 7$^{th}$, 2000,  corresponding to the first 
campaign; a second one with 10 observing nights between October 23$^{rd}$ and December 
13$^{th}$ 2001; and the third campaign which consisted of 71 nights from May 24$^{th}$, 2001 
to May 3$^{rd}$, 2002. We observed bright stars, almost always $m_{V}\lesssim 2.5$.

The high-altitude and precarious initial development of the site made the initial runs a 
real challenge. Nevertheless we successfully carried out measurements,
for the first time, during the night at the summit.  
To compare all measurements, we present  statistics giving equal weight to 
every data point.  
For the analysis we only consider data files with at least 20 points acquired
close to the zenith (airmass$\leq 1.15$) with non-saturated images (DIMM parameter 
pixmax $\leq 255$) to ensure a reliable stellar centroid determination.

\subsection{Results on seeing statistics}
\begin{figure}
\plotone{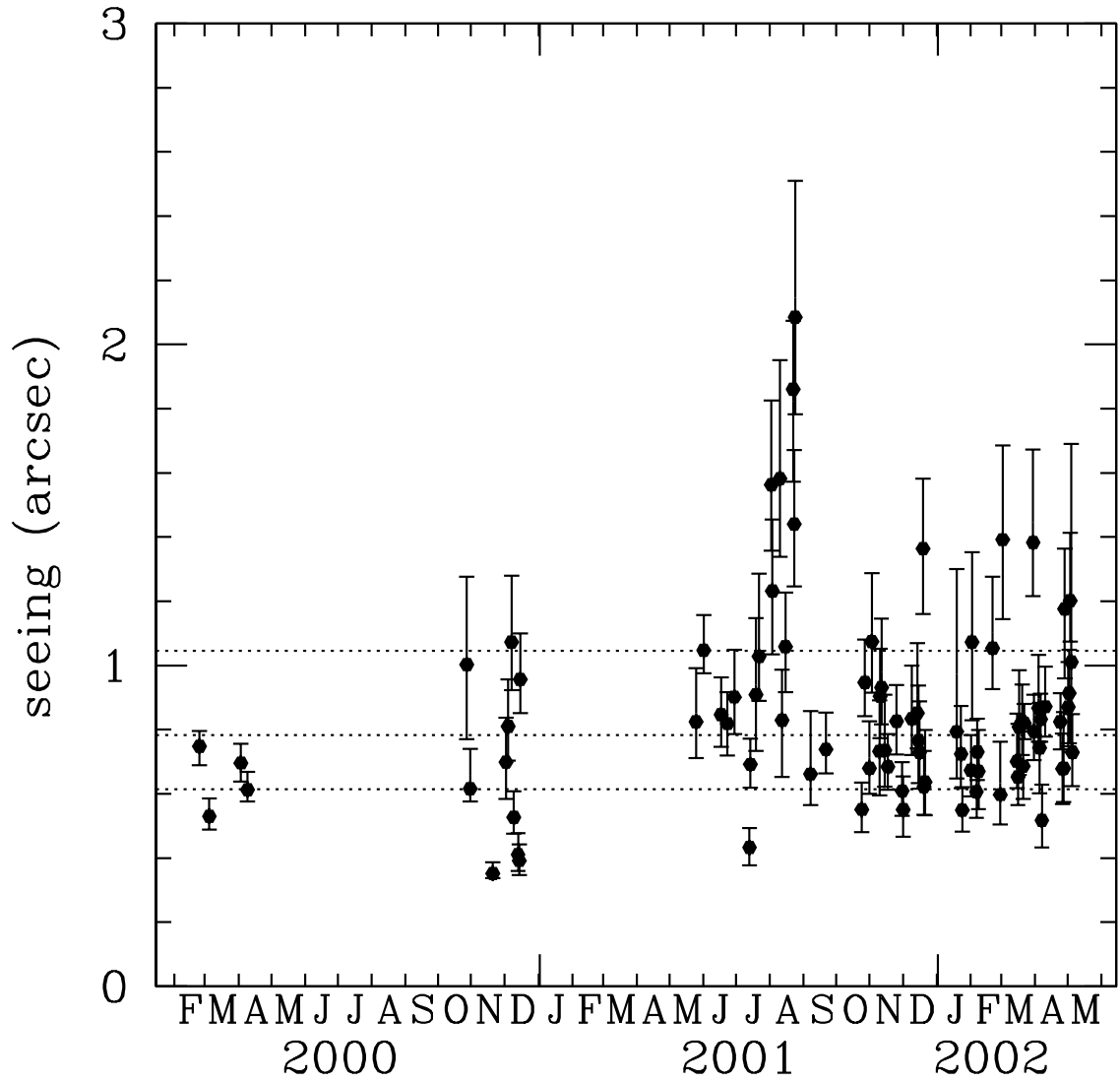} 
\caption{Seeing daily statistics for all data points. The dots are the median and the error
bars go from the first to the third quartile. The central dotted line denotes the
seeing median for the whole data set, corresponding to {\bf 0.78}". The top dotted line is
the distribution third quartile, equal to 1.04", and the bottom dotted line is the
first quartile, corresponding to 0.61".\label{s_daily}}
\end{figure}

\begin{figure}
\plotone{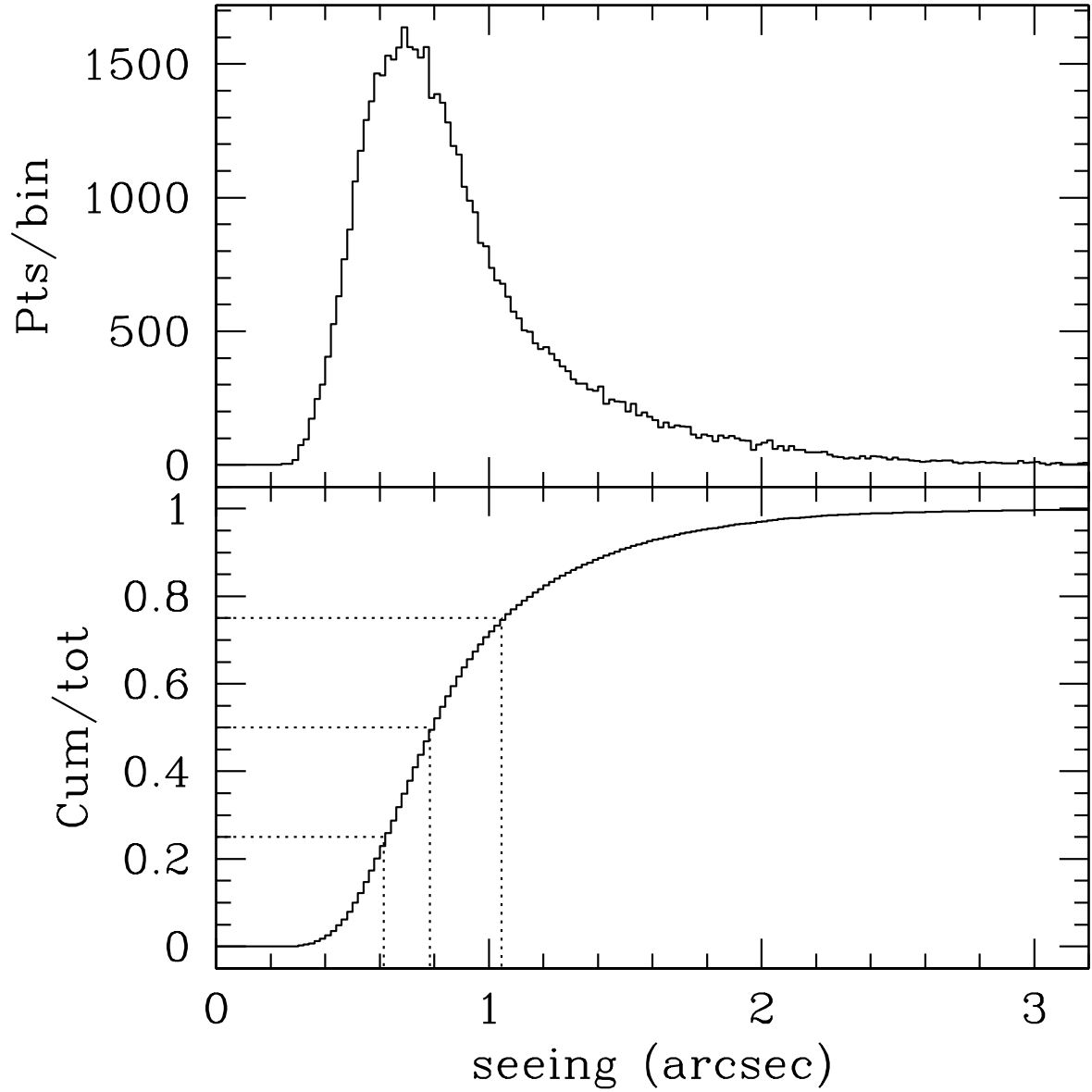}
\caption{Seeing histogram ({\em top}) and cumulative ({\em bottom}) distribution of all data 
points. The global 
statistics show that for 85 observing nights spanning from February 2000 to May 2002. 
Seeing values of 0.61" were achieved  25\% of the time, seeing below 0.78" 
during 50\% of the time and sub-arcsec seeing 70\% of the time.\label{s_distr_tot}}
\end{figure}

%------------------------------------------%
Figure~\ref{s_daily} is the daily plot of all the measurements. The dots are the median and 
the error bars go from the first to the third  quartile. The central dotted line denotes the
seeing median for the all data set, corresponding to 0.78". The top dotted line is
the distribution third quartile, 1.05" and the bottom dotted line is  the first quartile, 
corresponding to 0.62". The histogram and the cumulative distribution of the 
same data are shown in figure \ref{s_distr_tot}. It can be appreciated a
sub-arcsec seeing 75\% of the time.

\begin{figure}
\plotone{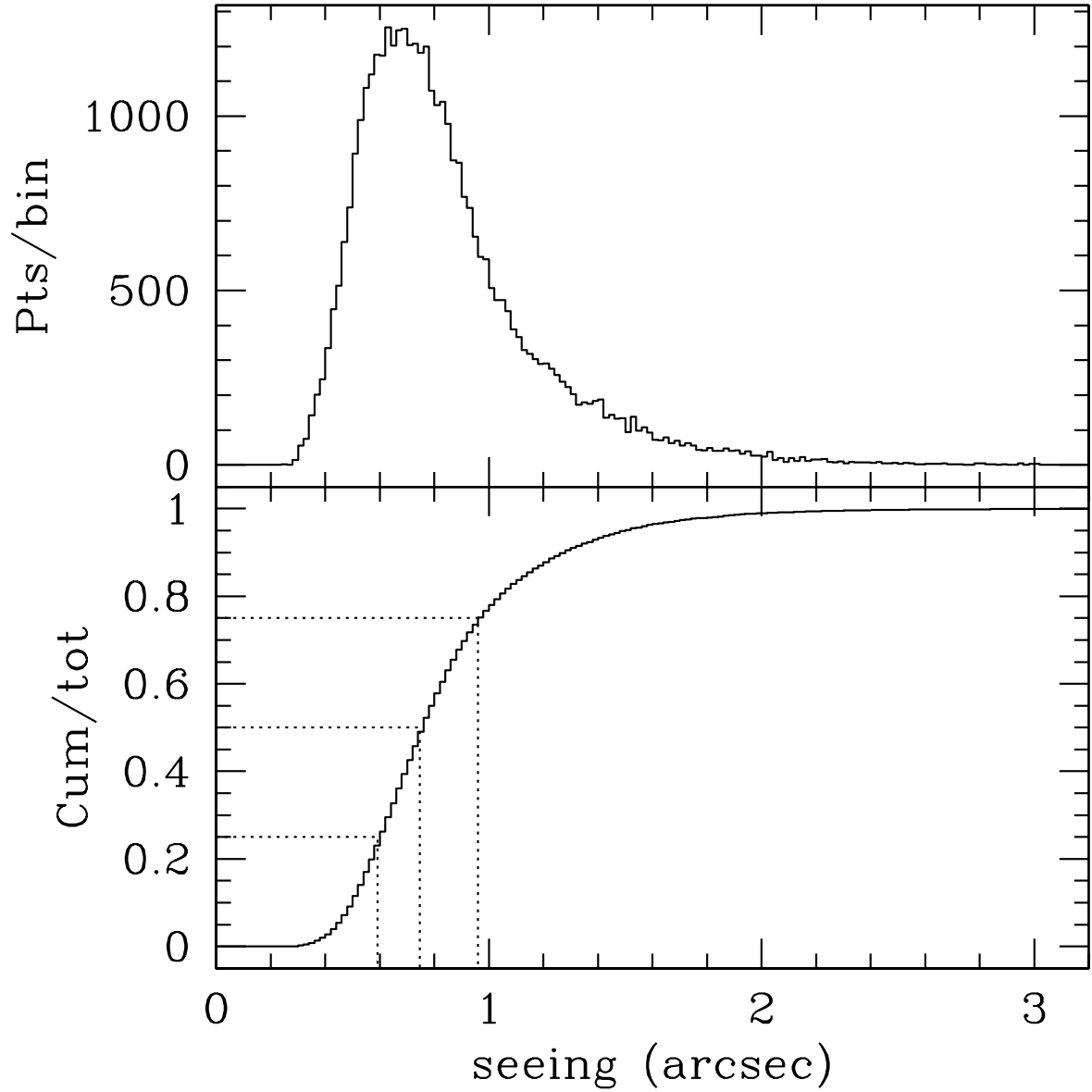}
\caption{Histogram ({\em top}) and cumulative ({\em bottom}) seeing distribution for the {\bf dry season} 
{\em i.e.} from November to April. The seeing median, indicated by the central 
dotted line, is {\bf 0.75}". The distribution first quartile, equal to 0.59", 
and the third quartile value, equal to 0.96", are indicated by the bottom and 
top dotted lines respectively. \label{dry_season}} 
\end{figure}

\begin{figure}
\plotone{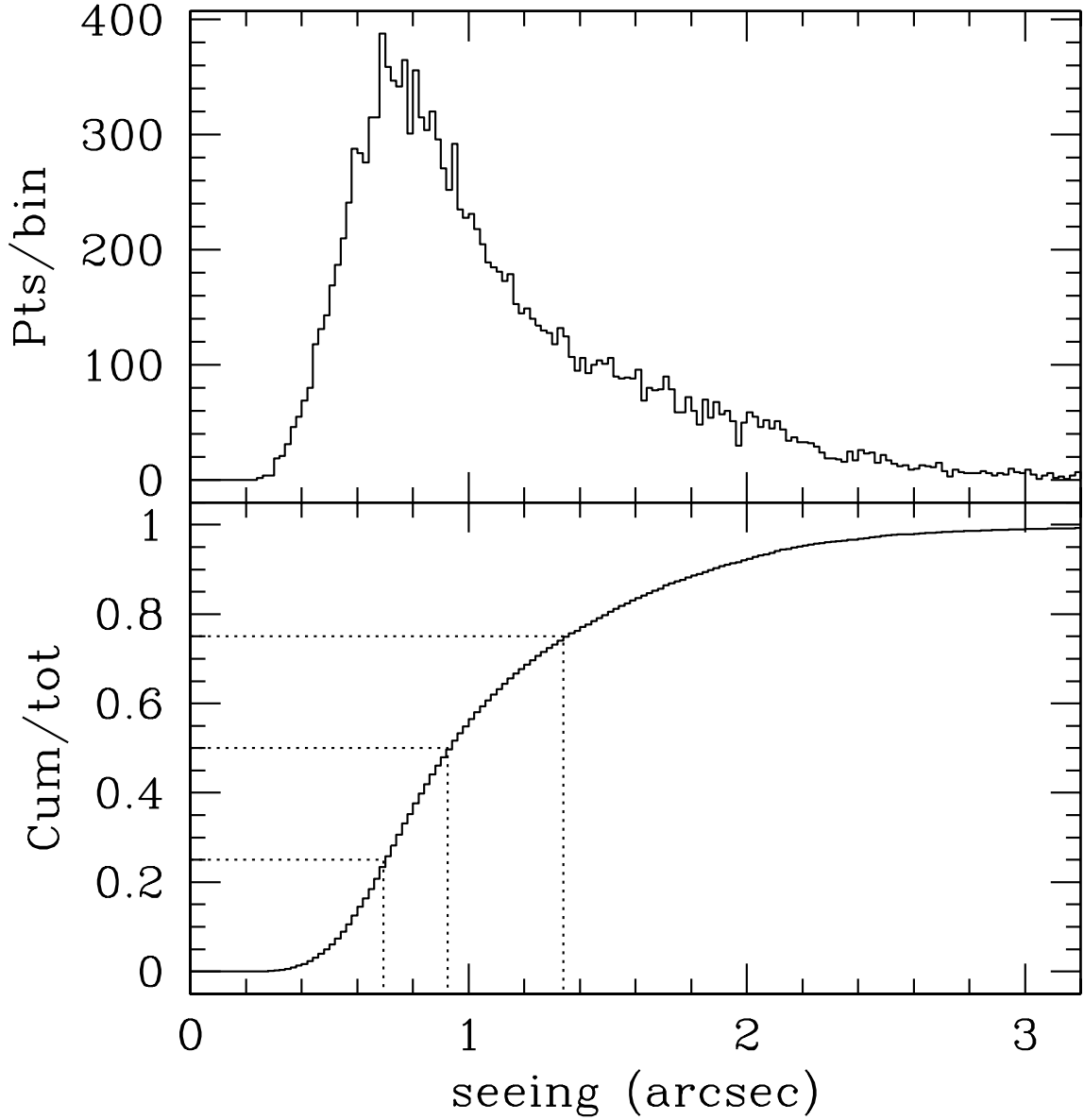}
\caption{Histogram ({\em top}) and cumulative ({\em bottom}) seeing distribution for 
the {\bf wet season}, 
{\it i.e.} from May to October. The seeing median, indicated by the central dotted 
line is {\bf 0.92}". The bottom dotted line corresponds to the first quartile value, 
equal to 0.69", and the third quartile value, equal to 1.34", is indicated by the top 
dotted line.\label{wet_season} }  
\end{figure}

To investigate the seeing seasonal behaviour we define the dry season from November to
April and the wet season from May to October. The histogram and cumulative distribution
for the dry and the wet seasons are shown in figures ~\ref{dry_season} and 
\ref{wet_season} respectively. While the seeing median during the dry season is 0.75",
for the wet season it raises 0.92". However it  should be noticed that during August 2001 
the seeing was specially bad (with a median of 1.49"; see table~\ref{table1}). To study 
the August contribution to the seeing we made the exercise of calculating statistics 
for the wet season without August 2001, the seeing median becoming 0.78", significantly 
closer to the dry season seeing median.

To compare the monthly seeing behaviour we present, in table~\ref{table1}, statistics giving 
equal weight to each individual seeing measurement for the complete data set: the first column 
is the month, the next
three columns give the details of the data acquired {\em i.e.} number of nights, number of
observing  hours during those nights and number of points. The next columns give the data 
statistics: mean, standard deviation, minumuum value, first quartile, median and third 
quartile. The global statistics shows that for 85 observing nights spanning from 
February 2000 to May 2002, the  seeing median is 0.78" with a standard deviation of 0.44". 

\begin{figure}
\plotone{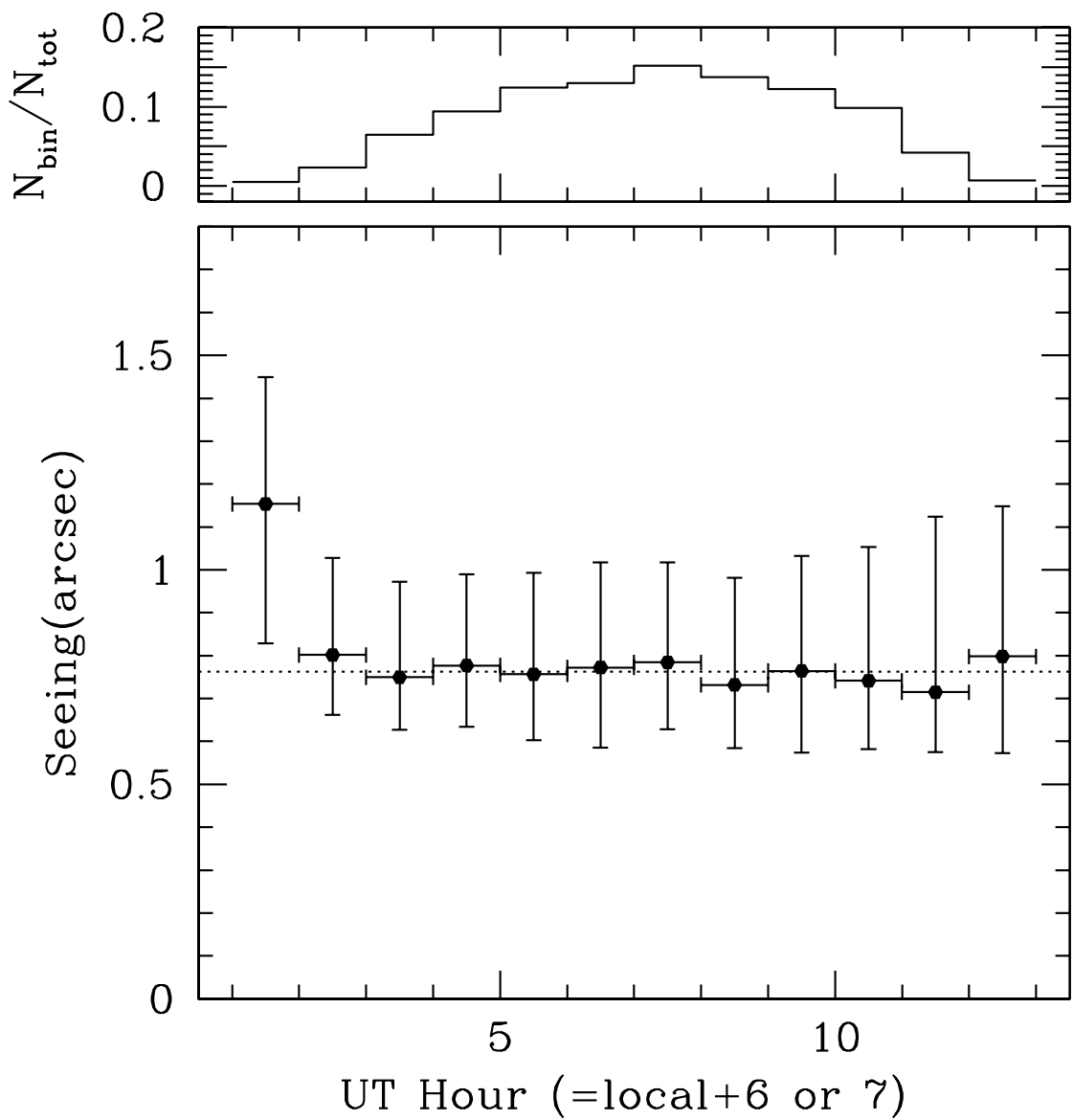}
\caption{Distribution of seeing values for different hours of the night
(in UTC). The dots mark the median values while the error bars go from the
first to the third quartile. The upper panel shows the number of data points
for each hour interval. Note that the 1-2 hrs UTC bin (7-8:pm local) has
a much lower coverage than the middle of the night, so the deviation of 
uniformity for that bin might be apparent and due to low number statistics
({\em i.e.} dominated by one or two nights with bad seeing). The data are basically 
consistent with no systematic trend during the night.\label{shourly}}
\end{figure}

In figure~\ref{shourly} the distribution of seeing values as a function of  UT is shown for 
the 85 observing nights. The dots represents the median for each hour and the error bars go 
from the first to the third  quartile. The histogram is shown in the upper panel. It must the 
noted that the first bin, corresponding to 7-8 PM local time, has only a few points so the 
high seeing value might be due to low number statistics rather than a intrinsically higher 
seeing at the beginning of the night. We conclude that our data do not show any systematic 
trend along the night. \cite{mvv97} observed that there is no general trend in the seeing 
evolution for the Roque de los Muchachos Observatory. In contrast~\cite{gio01}, point out 
%% begins comment 2:
that for high altitude cordillera sites, in northern Chile,  the seeing  tends to be of lower
quality at the beginning of  the evening.
%% ends comment 2:

\subsection{The seeing integration time}
%------------------------------------------%
It has been discussed by several authors that the temporal averaging of the variance of the 
differential motion with a finite exposure time depends on the average velocity and the 
direction of displacement of
the wavefront corrugation with respect to the DIMM aperture~\citep{ma87,zt97}.  
\citet{gio01}, on high altitude  cordillera sites in Chile, measure the seeing 
alternating 10~msec and 20~msec exposures. 
They obtain seeing estimates derived from 10~msec exposures, from 20~msec exposures and extrapolations to 
``zero exposure", obtained by multiplying the 10~msec seeing by the ratio of the 10 and  20~msec measurements. 
They find that the median values of the 0 ms seeing vary between 0.66" and 0.76", those of the 10~msec seeing 
between 0.56" and 0.65" while those of the 20~msec between 0.48" and 0.56". According to these
authors, the seeing for the 10~msec series is statistically worse than that for the 20~msec 
series, as the latter smears the image motion somewhat.

\begin{figure}
\plotone{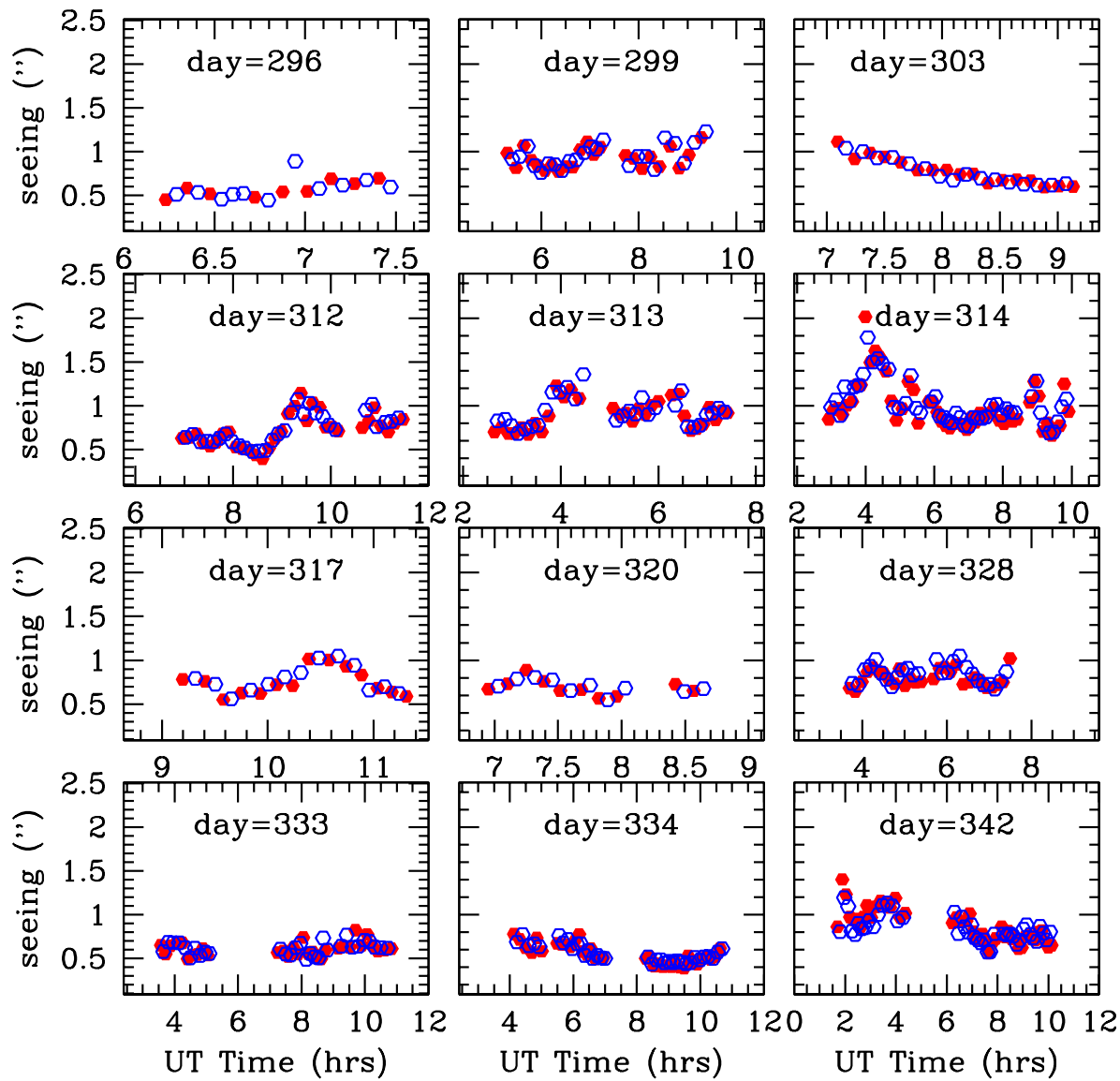}
\caption{Seeing time profiles for 12 nights with samples of 15 seeing measurements
taken alternating with 10~msec and 20~msec integration times. Each point in the figure 
represents the median over each sample, 
distinguishing between 10~msec samples (filled circles) and 20~msec 
samples (open circles). There is no clear qualitative systematic difference
between both data, all taken in 2001.\label{alternados_1}}
\end{figure}

\begin{figure}
\plotone{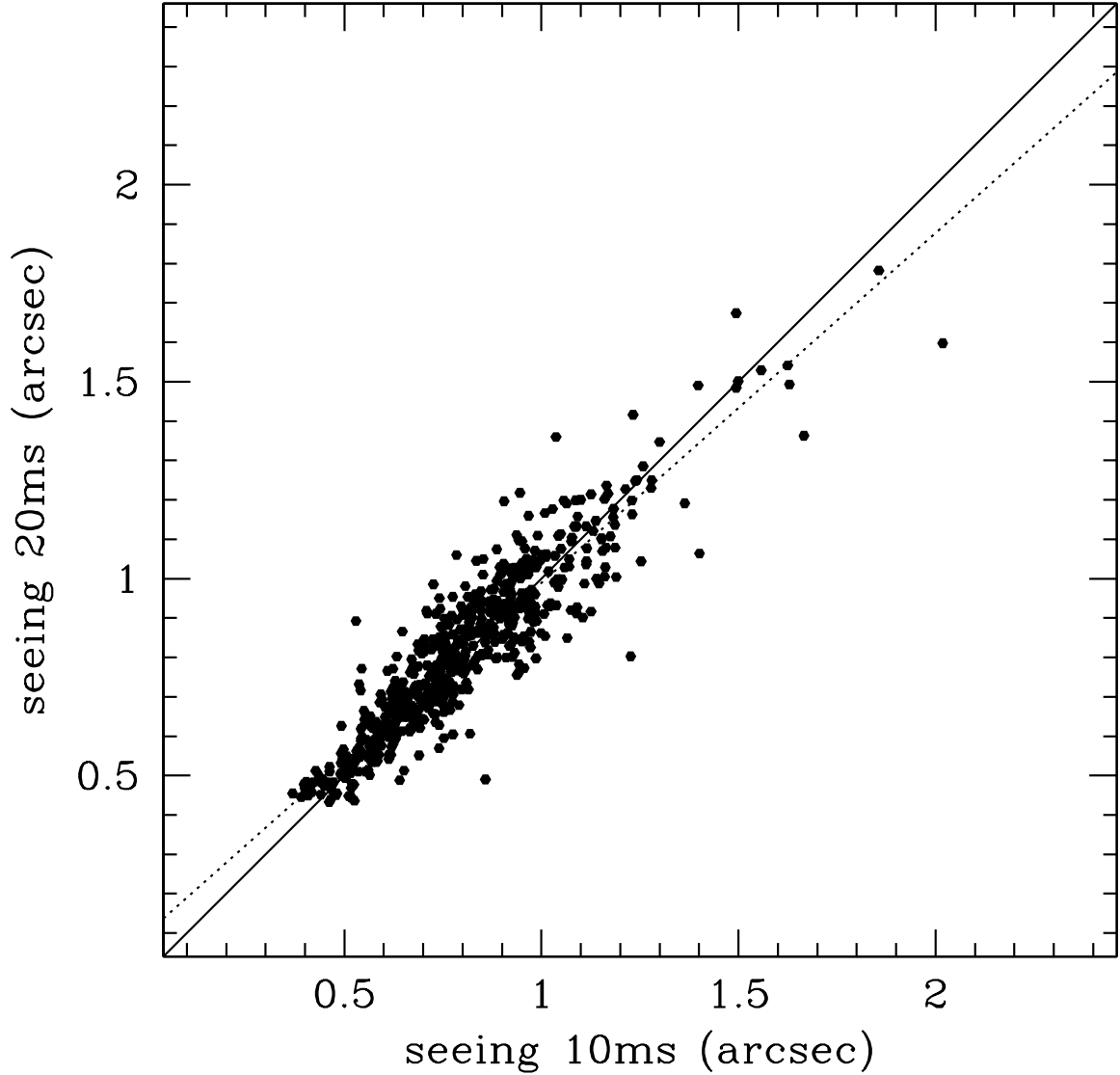}
\caption{Comparison of seeing data taken with integration times of 10 and 20~msec.
Each data set has been extended through a spline fit to overlap in time
the other data set, so that each point represents one sample (either 10~msec
or 20~msec) and the reciprocal spline interpolation (from the 20~msec or
10~msec data). The full line represents $s_{10}=s_{20}$ while the dotted
line is the best fit to the data: $s_{20} = (0.888\pm 0.205)s_{10} + (0.102\pm 0.152)$,
with a correlation of 0.92 and rms=0.08". Errors were determined with the bootstrap method. 
\label{alternados2}}
\end{figure}

Our DIMM has a default integration time of 20~msec. The  first data set that spans from 
February 2000 to August 2001 were taken using that integration time. 
During the next 12 observing nights we alternated measurements with 10 and 20~msec 
integration times. The camera control allows the user to alternate between the default time 
and a mechanically selected integration time. As the selection is manual we decided to take 
15~integrations at each integration time. The results are shown in Figure \ref{alternados_1} 
where the seeing as a function of time is shown for each night. The filled circles 
corresponds to the medians of the 10~msec integration samples while the open circles to 
the medians of the 20~msec samples. Qualitatively, there is no significant difference between 
the trends of both data sets. We compare quantitatively the samples by plotting the 10~msec 
seeing medians {\em vs.} 20~msec seeing medians where each data set has been interpolated
through a spline fit to overlap in time the other data set, such that at each time we
have one data point (either from the 10~msec or 20~msec sample) and one spline interpolation
(from the 20~msec or 10~msec sample) which can be compared. The comparison is shown in 
figure~\ref{alternados2}, where the best fit to the data is the dotted line and the full 
line represents $s_{10}=s_{20}$.  The best fit slope $(0.89\pm 0.20)$ and intercept 
$(0.10\pm 0.15)$ are compatible with $s_{10}=s_{20}$.

\begin{figure}
\plotone{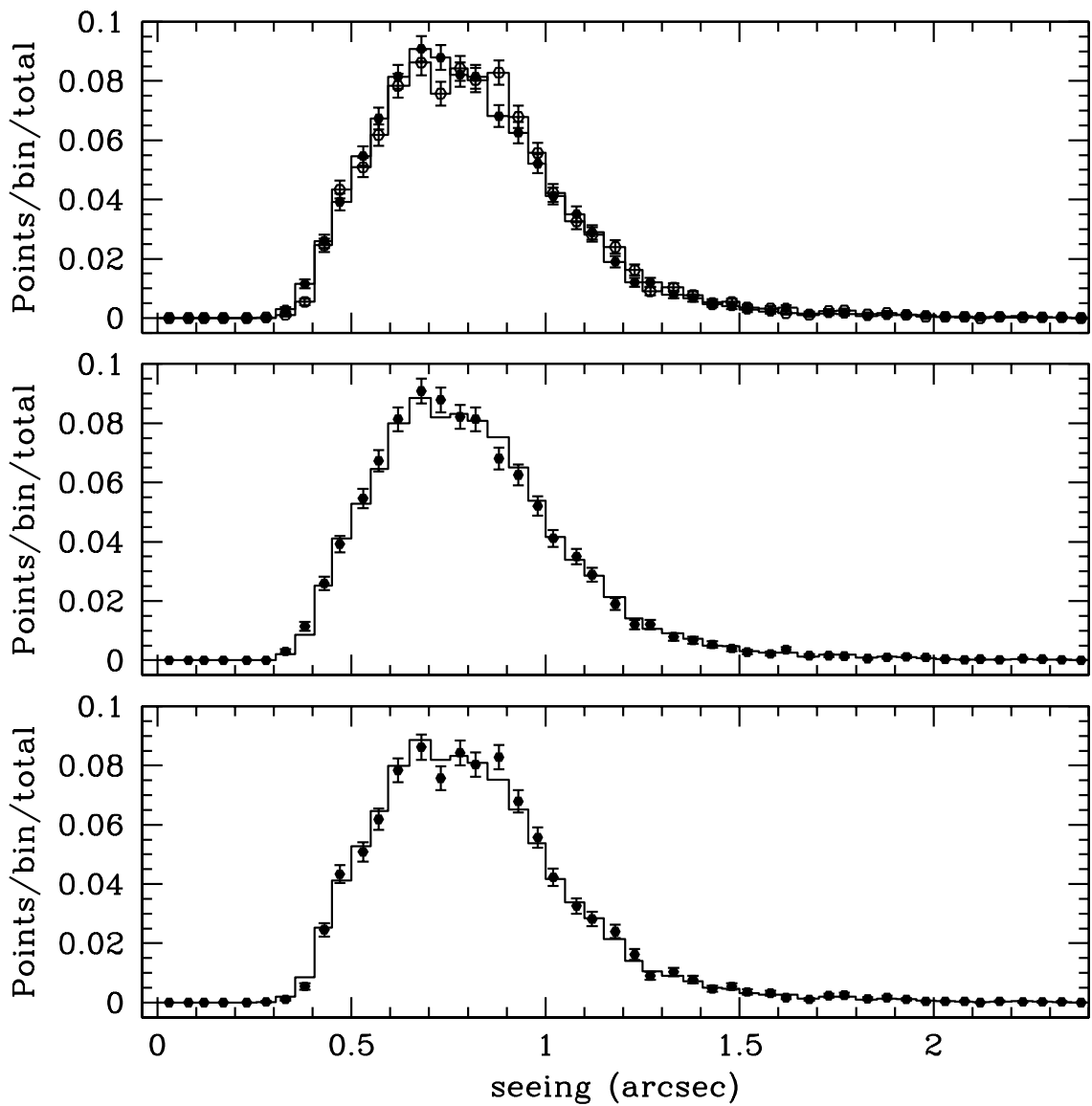}
\caption{Comparison between the distributions of $s_{10}$ and $s_{20}$ values ({\em top});
between the common distribution and $s_{10}$ values ({\em middle}); and between the common 
distribution and $s_{20}$ values ({\em bottom}).
As expected from the figure, the quartile and median values are similar. 
In particular $q_{2}(s_{10}) = 0.76"$, $q_{2}(s_{20}) = 0.78"$ and $q_{2}(s_{10}\& s_{20}) = 
0.77"$. Both $s_{10}$ and $s_{20}$ distributions are compatible with the same single parent 
distribution. \label{alternados_distr}}
\end{figure}

We also compared the complete distributions of 10~ms and 20~ms integration times seeing 
values, shown in figure~\ref{alternados_distr}. A $\chi^{2}$ comparison test between both distribution
gives $\chi^{2} = 64.8$ for 46 degrees of freedom, that is a 3.5\% probability that both 
distributions are the same. However, if we compare both distributions with the common
distribution, derived from putting together the samples, then the respective $\chi^{2}$
values are 10.1 and 12.1 for the same number of degrees of freedom, giving respective 
probabilities of $1-(4\times 10^{-9})$ and $1-(10^{-7})$ that each distribution can be derived
from the same parent distribution. As it can be appreciated in the two lower panels of 
figure~\ref{alternados_distr}, the two distributions of seeing values are compatible with a 
single parent distribution. The total seeing median including both integration times is 0.77".

%%%  Begins comment 3

\begin{figure}
\plotone{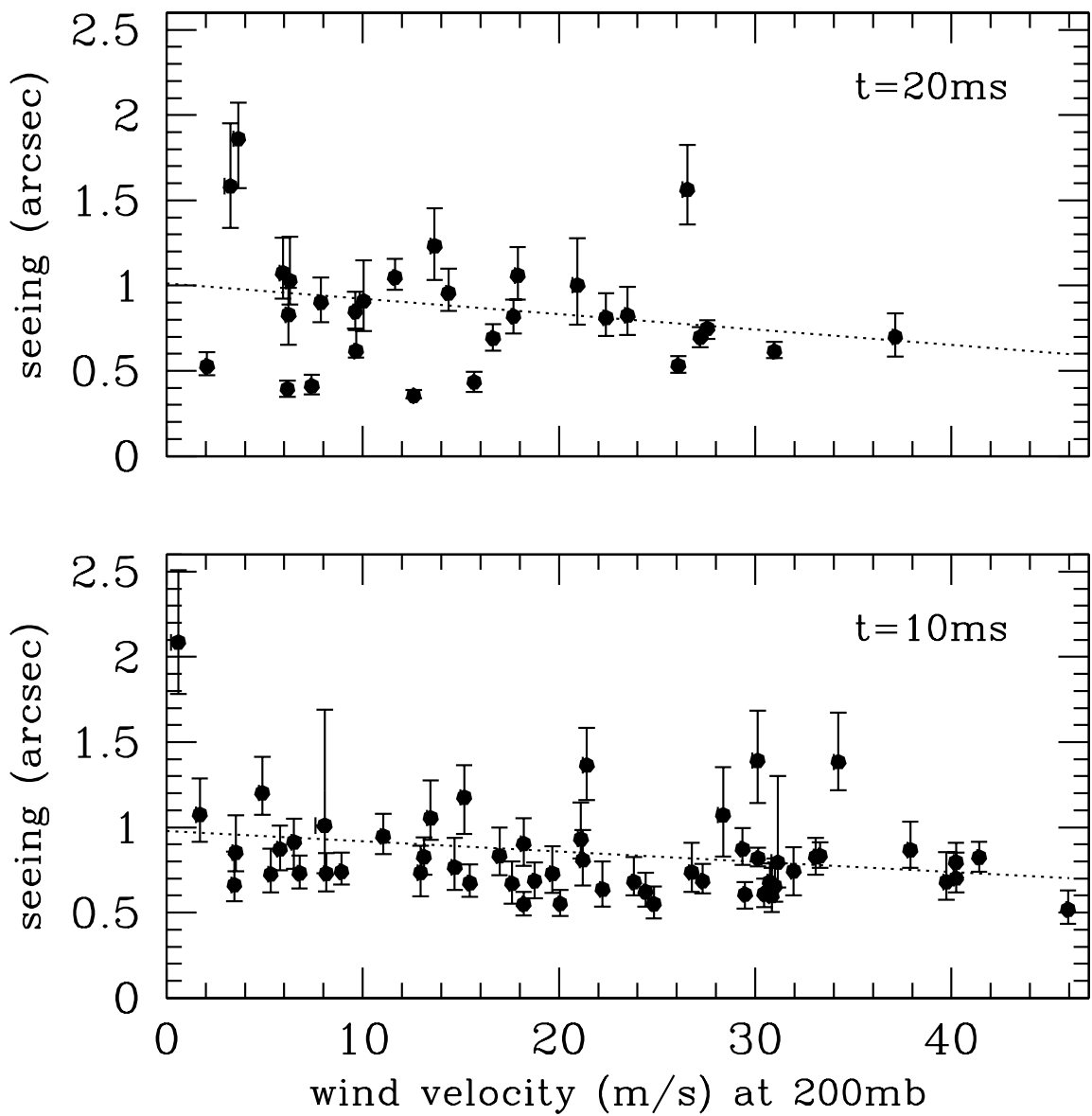}
\caption{Seeing median as a function of the wind velocity at 200~mb.  The seeing error bars go 
from the first to the third quartile.  The daily average wind velocities were obtained from the 
NOAA Global Gridded Upper Air data base. Seeing data for the 20 ms sample ({\em top}): the dotted 
line represents the best linear fit, of slope $=-0.009\pm 0.008$, consistent with a zero slope 
within 1.1$\sigma$. The correlation coefficient is equal to $-0.274$ and a rms dispersion is 0.260". 
Seeing data for the 10~ms sample ({\em bottom}): the best linear fit, slope~$=-0.006\pm0.004$, 
is consistent with a slope equal to zero within 1.5$\sigma$. The correlation coefficient is $-0.227$ 
and the rms dispersion 0.359". 
\label{seeing_200mb}}
\end{figure}

Following a suggestion by Marc Sarazin (private communication), we studied the presence of 
temporal averaging effects on our seeing measurements by comparing the seeing median {\em vs.} the wind 
velocity at 200~mb using the NOAA Global Gridded Upper Air data base, for each night of our 20~msec sample. 
The results are shown in the upper panel of Figure \ref{seeing_200mb}. The seeing error bars  
go from the first to the third quartile. The wind data are daily  average of four measurements available 
on the NOAA database. 
The dotted line represents the best least square linear fit, consistent with slope equal to zero 
within 1.1$\sigma$. 
The correlation coefficient is equal to $-0.227$ and the rms dispersion is 0.359". The errors
were obtained using a bootstrap technique. The data show that the seeing does not drop at high speed 
as it would be expected in the presence of temporal averaging effects. In the lower panel of the same 
figure, the 10~ms seeing data are shown. In this case the best fit is also consistent with zero slope, 
within 1.5$\sigma$, and the correlation coefficient is $-0.274$.

\begin{figure}
\plotone{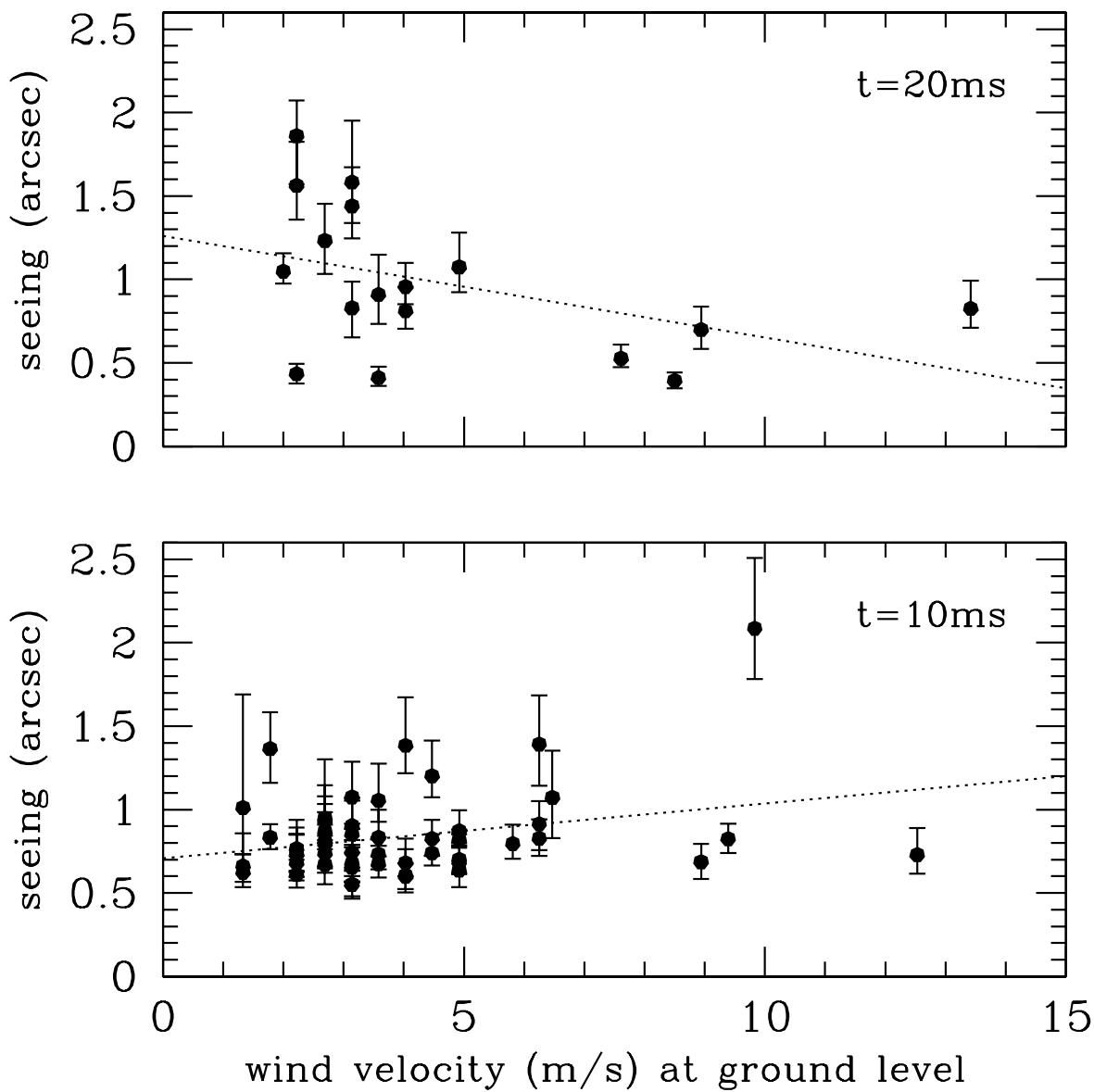}
\caption{Seeing median as a function of the wind velocity median at ground level for those nights that
have both seeing and wind velocity data. The seeing error bars go from the first to the third quartile, 
while the wind velocity error bars are not included for clarity. Seeing data for 17 nights of the 20~ms 
sample ({\em top}): the dotted line represents the best linear fit, slope $=-0.06\pm 0.04$, which is 
consistent with a null slope within 1.4$\sigma$.
The correlation coefficient is $-0.435$ and a rms dispersion 0.384". Seeing data for 50 nights of 
the 10~ms sample ({\em bottom}): the best fit, slope $=0.03\pm 0.03$, is also consistent with a slope $=$ 0 
within 1.1$\sigma$. The correlation coefficient is 0.281 and the rms dispersion 0.256". 
\label{seeing_ground}}
\end{figure}

To study the presence of any correlation between  seeing and wind velocity at ground level, we
compare the seeing and  wind velocity daily medians at the site.
The wind velocity was measured with a meteorological station located on the seeing monitor tower.  
Figure~\ref{seeing_ground} shows the seeing median as a function of the wind velocity median 
for those nights that have simultaneous seeing and wind velocity data. 
The seeing error  bars  go from the first to the third quartile.  The wind velocity 
distribution median was calculated from the data obtained between 8:pm and 6:am local
time, the error bars corresponding to the first and third quartile are not included in the plot for clarity. 
In the upper panel the results for 17 nights of the 20~msec sample are shown. 
The dotted line represents the best fit that is consistent with zero slope within 1.4$\sigma$, a 
correlation factor equal to $-0.435$ and a rms dispersion of 0.384".  In the lower panel the data obtained for
50 nights of the 10~ms sample are shown.  The best fit is consistent with slope equal to zero 
within 1.1$\sigma$, a correlation factor of  0.281 and a rms dispersion equal to 0.256".

%%% Ends comment 3

In contrast with 
\citet{gio01} results on high altitude cordillera sites in  Chile, the data for Sierra Negra 
suggest that, within our statistics, there is no difference between the 10~msec and the
 20~msec series. It must 
be noted that \cite{vm95} find the seeing bias produced by the difference in exposure
time to be highly dependent on the magnitude of the star. As we generally use stars brighter
than $m_{V} = 2$, it is possible that this bias is in our data but below measurable error,
therefore not influencing our results. Nevertheless from December 2001 onwards all our 
measurements are made using 10~msec exposures. 

%%%%%%%%%%%%%%%%%%%%%%
\section{Conclusions}
%%%%%%%%%%%%%%%%%%%%%%
We present the first seeing measurements carried out at Sierra Negra. For 85 observing nights, 
seeing values of 0.62" were achieved 25\% of the time, below 0.78" during 50\% of the 
time, and sub-arcsec seeing for 70\% of the time.  The comparison  between the dry  
and the wet season shows that the seeing median is better during the dry season with a value
of 0.73" and sub-arcsec  seeing 77\% of the time. 
In contrast for the wet season the seeing 
median is 0.92" value is strongly affected  by  August 2001 contribution.
We analysed the dependence of the seeing statistics with time of night without 
finding any systematic trend of seeing as a function of time.
We did not find any correlation between the seeing values and the 200~mb wind velocity. A preliminary
analysis of the correlation between the seeing and  the wind velocity at ground level was carried out. The
 results show that there is not an  obvious correlation between them. 
Nevertheless we will continue analysing  the seeing as a function
of other meteorological parameters in more detail  to try to find  out  where  most of the turbulence
is concentrated.

The results obtained so far show that Sierra Negra is a competitive site for optical 
astronomy. We will continue with our seeing and meteorological measurements to characterise 
the site on a longer time scale basis. We have carried out an independent analysis of the 
seeing temporal structure and the results will be reported in a separated  paper.

%%%%%%%%%%%%%%%%%%%%%%%%%%%%%%%%%

\acknowledgments {We thank R.J. Terlevich his interest and his useful comments and suggestions.  This study was  
possible through funds from the Instituto Nacional de Astrof\'{\i}sica, Optica y Electr\'onica. 
The 20mb wind velocity data was provided by the NOAA-CIRES Climate Diagnostics Center, Boulder, Colorado,
USA, from their Web site at http://www.cdc.noaa.gov/.}

%%%%%%%%%%%%%%%%%%%%%%%%%%%%%%%%%%

%%%%%%%%%%%%%%%%%%%%%%%%%%%%%%%%%%
 
\clearpage

\clearpage

\clearpage

\clearpage

\clearpage

\clearpage

\clearpage

\clearpage

\clearpage

\clearpage

\clearpage

\clearpage

\begin{deluxetable}{lrrrrrrrrr}
\tabletypesize{\small}
\tablecaption{Global statistics per month considering equal weight per each seeing datum.
\label{table1}}
\tablewidth{0pt}
\tablehead{
\colhead{Date}  & \multicolumn{3}{l}{Data acquired} & \multicolumn{6}{l}{Data statistics (")}\\
\colhead{Month Year} & \colhead{Nights} & \colhead{Hours} & \colhead{Points} & \colhead{Mean} &
\colhead{$\sigma$} & \colhead{Min} & \colhead{$q_{1}$} & \colhead{Median} & 
\colhead{$q_{3}$}
}\startdata
 Feb 2000 &  1 & 0.3  &   42 & 0.756 & 0.087 & 0.606 & 0.689 & {\bf 0.748} & 0.795  \\
 Mar 2000 &  1 & 1.2  &  227 & 0.541 & 0.077 & 0.397 & 0.488 & {\bf 0.531} & 0.586 \\ 
 Apr 2000 &  2 & 1.5  &  205 & 0.650 & 0.132 & 0.482 & 0.578 & {\bf 0.631} & 0.696 \\
 Oct 2000 &  2 & 2.3  &  389 & 1.035 & 0.344 & 0.460 & 0.744 & {\bf 0.958} & 1.259 \\
 Nov 2000 &  2 & 3.5  &  427 & 0.701 & 0.204 & 0.290 & 0.553 & {\bf 0.678} & 0.824 \\
 Dec 2000 &  6 & 25.5 & 2383 & 0.661 & 0.354 & 0.238 & 0.401 & {\bf 0.506} & 0.894 \\
 May 2001 &  2 & 1.7  &  302 & 0.887 & 0.204 & 0.552 & 0.722 & {\bf 0.854} & 1.021 \\
 Jun 2001 &  3 & 7.5  & 1178 & 0.910 & 0.468 & 0.466 & 0.749 & {\bf 0.848} & 0.965 \\
 Jul 2001 &  4 & 12.0 & 1926 & 0.736 & 0.374 & 0.229 & 0.488 & {\bf 0.670} & 0.870 \\
 Aug 2001 &  8 & 26.7 & 4467 & 1.589 & 0.602 & 0.476 & 1.172 & {\bf 1.489} & 1.894 \\
 Sep 2001 &  2 & 6.0  & 1063 & 0.783 & 0.307 & 0.329 & 0.576 & {\bf 0.677} & 0.857 \\
 Oct 2001 &  3 & 15.2 & 2009 & 0.811 & 0.258 & 0.376 & 0.616 & {\bf 0.789} & 0.962 \\
 Nov 2001 &  9 & 37.6 & 7352 & 0.826 & 0.312 & 0.290 & 0.622 & {\bf 0.776} & 0.959 \\
 Dec 2001 &  7 & 37.5 & 6380 & 0.835 & 0.326 & 0.343 & 0.614 & {\bf 0.755} & 0.963 \\
 Jan 2002 &  4 & 29.6 & 3837 & 0.782 & 0.342 & 0.326 & 0.574 & {\bf 0.679} & 0.840 \\
 Feb 2002 &  6 & 36.0 & 4207 & 0.861 & 0.404 & 0.272 & 0.602 & {\bf 0.758} & 1.002 \\
 Mar 2002 &  9 & 36.1 & 4867 & 0.980 & 0.431 & 0.293 & 0.669 & {\bf 0.846} & 1.203 \\
 Apr 2002 & 10 & 44.1 & 7616 & 0.788 & 0.284 & 0.274 & 0.588 & {\bf 0.742} & 0.913 \\
 May 2002 &  4 & 17.9 & 3202 & 0.972 & 0.447 & 0.347 & 0.675 & {\bf 0.834} & 1.100\\
\tableline
All data  & 85 &342.1& 52079 & 0.898 & 0.441 & 0.229 & 0.615 & {\bf 0.784} & 1.046 \\
\enddata
%\tablenotetext{}{All statistical values in arcsecs.}
\end{deluxetable}

\end{document}